\documentclass[sn-mathphys,Numbered]{sn-jnl}

\usepackage{graphicx}%
\usepackage{multirow}%
\usepackage{amsmath,amssymb,amsfonts}%
\usepackage{amsthm}%
\usepackage{mathrsfs}%
\usepackage[title]{appendix}%
\usepackage{xcolor}%
\usepackage{textcomp}%
\usepackage{manyfoot}%
\usepackage{booktabs}%
\usepackage{algorithm}%
\usepackage{algorithmicx}%
\usepackage{algpseudocode}%
\usepackage{listings}%
\usepackage{bm}

\newcommand{\bxi}{{\boldsymbol \xi}}
\newcommand{\bmu}{{\boldsymbol \mu}}
\newcommand{\bn}{{\boldsymbol n}}
\newcommand{\br}{{\boldsymbol r}}
\newcommand{\btheta}{{\boldsymbol \theta}}
\newcommand{\bpi}{{\boldsymbol \pi}}
\newcommand{\cL}{{\cal L}}

\newcommand{\pd}[2]{\frac{\partial #1}{\partial #2}}
\newcommand{\cov}{{\rm Cov}}

\begin{document}

\title[]{Fundamental precision limits of fluorescence microscopy: a new perspective on MINFLUX}

\author*[1]{\fnm{Matteo} \sur{Rosati}}\email{matteo.rosati@uniroma3.it}

\author[1,2]{\fnm{Miranda} \sur{Parisi}}

\author[3]{\fnm{Ilaria} \sur{Gianani}}

\author[3,4]{\fnm{Marco} \sur{Barbieri}}

\author[1]{\fnm{Gabriella} \sur{Cincotti}}

\affil[1]{\orgdiv{Dipartimento di Ingegneria Civile, Informatica e delle Tecnologie Aeronautiche}, \orgname{Universit\'a Roma Tre}, \orgaddress{\street{Via Vito Volterra 62}, \city{Rome}, \postcode{I-00146}, \country{Italy}}}

\affil[2]{\orgname{Universitas Mercatorum}, \orgaddress{\street{Piazza Mattei 10}, \city{Roma}, \postcode{I-00186}, \country{Italy}}}

\affil[3]{\orgdiv{Dipartimento di Scienze}, \orgname{Universit\'a Roma Tre}, \orgaddress{\street{Via Vito Volterra 62}, \city{Rome}, \postcode{I-00146}, \country{Italy}}}

\affil[4]{ \orgname{Istituto Nazionale di Ottica - CNR}, \orgaddress{\street{Largo E. Fermi 6}, \city{Firenze}, \postcode{I-50125}, \country{Italy}}}

\abstract{In the past years, optical fluorescence microscopy (OFM) made steady progress towards increasing the localisation precision of fluorescent emitters in biological samples. The high precision achieved by these techniques has prompted new claims, whose rigorous validation is an outstanding problem.
For this purpose, local estimation theory (LET) has emerged as the most used mathematical tool. 

We establish a novel multi-parameter estimation framework that captures the full complexity of single-emitter localisation in an OFM experiment. Our framework relies on the fact that there are other unknown parameters alongside the emitter's coordinates, such as the average number of photons emitted (brightness), that are correlated to the emitter position, and affect the localisation precision. The increasing complexity of a multi-parameter approach allows for a more accountable assessment of the precision.   

We showcase our method with MINFLUX microscopy, the OFM approach that nowadays generates images with the best resolution. Introducing the brightness as an unknown parameter, we shed light on features that remain obscure in the conventional approach: the precision can be increased only by increasing the brightness, ({\it i.e.}, illumination power or exposition time), whereas decreasing the beam separation offers limited advantages. 

We demonstrate that the proposed framework is a solid and general method for the quantification of single-emitter localisation precision for any OFM approach on equal footing,  evaluating the localization precision of stimulated emission depletion (STED) microscopy and making a comparison with MINFLUX microscopy. 
}




\maketitle

\section{Introduction}
\label{sec1}
In the last 30 years, optical fluorescence microscopy (OFM) has ushered a revolution in the life science research~\cite{Hell2007,Hell2015,Schermelleh2019,Lelek2021,Parisi2023}, enabling scientists to visualize and investigate molecular interactions, cellular structures, and dynamic processes with unprecedented sensitivity and specificity, thus uncovering fundamental insights into biological systems. A large variety of super-resolution techniques have been developed, that are able to resolve objects beyond the Abbe diffraction limit~\cite{Hell1994,Schwartz2012,GattoMonticone2014,Tenne2019}, up to a localisation precision of the order of a few nanometers achieved by MINFLUX~\cite{Balzarotti2017,Gwosch2020,Zhan2022,Prakash2023}. In an OFM experiment, the objects of interest are labelled by  fluorophores, {\it i.e.} specific fluorescent molecules, that are excited by a light beam of suitable wavelength, and  radiatively decay, resulting in the emission of photons, typically with Poissonian statistics. 
OFM techniques differ mainly depending on the shape of the illuminating beam, the detection method, and the post-processing algorithm applied to the raw data. In structured illumination microscopy (SIM), sub-diffraction information can be extracted using a Fourier-transform based approach \cite{Heintzmann2017}. Image resolution can be augmented in stimulated emission depletion (STED) microscopy, employing an additional doughnut-shaped depletion laser \cite{Eggeling2015}. Stochastic optical reconstruction microscopy (STORM) and photo-activated localisation microscopy (PALM) randomly activate different fluorophores \cite{Henriques2011}.

Such a wide display of available methods, and the increasing resolution they aim to deliver require a rigorous and complete mathematical framework to quantify and compare the localisation precision of different OFM techniques. In this setting, local estimation theory (LET) has been used in recent literature~\cite{Chao2016,Balzarotti2017,Ober2020,Kalisvaart2022,Zhan2022,Xiu2023}, to estimate unknown parameters $\bxi = (\xi_1,\xi_2,\cdots,\xi_K)$, starting from an approximate value. Knowing the detection results $\bn = (n_1,\cdots,n_D)$, in the form of photon counts at different positions or instant times, as well as their statistics, LET is useful in two respects. First, it provides the Cram\'er-Rao bound (CRB), {\it i.e.} a lower bound on the minimum mean square error that an algorithm could attain trying to estimate $\bxi$, using the detection results $\bn$. Second, it provides an optimal estimator that actually saturates the CRB in the limit of a large number of repetitions of the experiment, as, for instance, the maximum-likelihood estimator (MLE).

Previous works have applied LET considering the fluorophore spatial coordinates $\bxi$ as the only unknown parameters~\cite{Chao2016,Balzarotti2017,Ober2020,Kalisvaart2022,Zhan2022,Xiu2023}. However, along with these target parameters,  other physical quantities generally govern the detection results. Although such additional parameters possess limited interest, and are commonly regarded as a nuisance, they need to be estimated nevertheless. To the best of our knowledge, previous works rely on the assumption that nuisance parameters are known (single-parameter analysis), or else they can be factored out of the statistical model. For instance, the number of photons emitted by a fluorophore has a mean value depending on factors other than its position. Hence, the localisation precision obtained considering the nuisance parameters fixed  may not accurately reflect the actual measured data.

\begin{figure}[H]
    \centering
    \includegraphics[width=\textwidth]{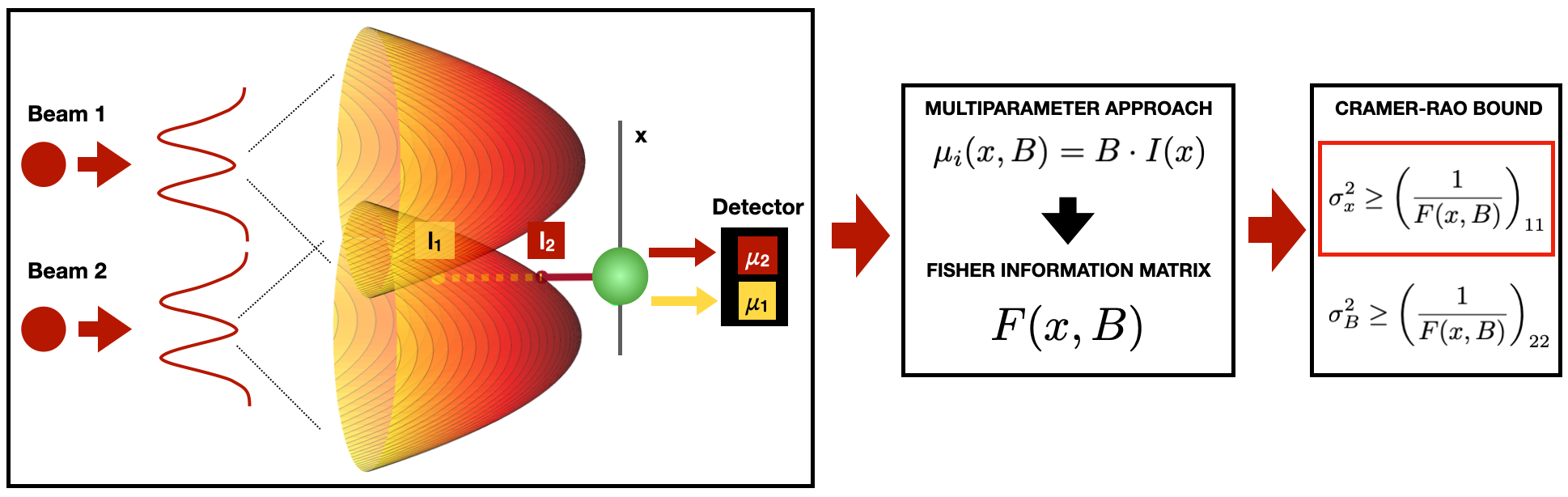}
    \caption{Conceptual scheme of multi-parameter estimation. Based on  the experimental technique (MINFLUX microscopy in this case), one identifies the parameters that describe the detection statistics. These include both the target parameters  (emitter position $x$) and additional nuisance parameters (effective brightness $B$). The minimum MSE of the estimation is  a function of all parameters, and it can be computed via the Fisher information matrix . }
    \label{fig:method}
\end{figure}

In light of these considerations, in this article we call upon an accurate evaluation of the fluorophore locatization precision into a wider framework, based on multi-parameter estimation. This approach is schematically represented in Fig.~\ref{fig:method}, highlighting that  all unknown parameters in the experiments are properly taken into account, because their accurate estimation deeply affects the localisation precision. We establish such a framework in full generality, and then apply it to OFM techniques where the mean number of detected photons is a priori unknown, viz. MINFLUX and STED microscopy. We believe that this result fills a significant void in the literature: our methodology  introduces a standard for the localisation precision of different OFM techniques. 

In order to illustrate our approach, we start by referring to MINFLUX microscopy. This relies on multiple illuminating beams, separated by a distance $L$, presenting a dark spot of width $s$.
We analyse the localisation precision of MINFLUX microscopy under the lens of our method, and demonstrate that the minimum mean square error (MSE) $\sigma^2$ for the localisation of an emitter scales as
\begin{align}
    \sigma^2 \propto \frac{s^2}{B},
\end{align}
when the fluorophore is close to the middle point between the beams. 
Here $B$ is the effective brightness parameter controlling the mean number of photons measured during the experiment. 
Therefore, the estimation error predicted in the multi-parameter framework does not depend on the beam separation $L$, but only on the width of the dark spot  $s$ and the mean number of detected photons, {\it i.e.} illumination intensity or exposure time. We retrieve the previous estimate~\cite{Balzarotti2017} under specific conditions, and corroborate our prediction via extensive numerical simulations.

The introduction of nuisance parameters enables a rigorous comparison of different OFM techniques that, due to their inherent different implementations, are difficult to compare. Indeed, if one is able to identify all the nuisance parameters that characterize an OFM technique, the proposed multi-parameter framework can accurately predict the localisation precision, as a function of all parameters. In this way, different techniques are put on an equal footing, and it is possible to highlight the dominant role of some parameters in the  localisation precision. As an example, we compare the localisation precision attainable with MINFLUX microscopy and STED, using the same nuisance parameter (brightness $B$).

\section{Results}\label{sec:results}

\subsection{Emitter localisation as a multi-parameter estimation problem}\label{sec:general}
In a typical OFM experiment, the main goal is the  estimation of the spatial coordinates of one or more emitters in the sample. We call $\bxi$ the vector comprising the actual coordinates of all the $k$ emitters, {\it e.g.}, $\bxi = (x_1,y_1,z_1,\cdots,x_k,y_k,z_k)$. 
The detection results are represented by the vector $\bn = (n_1,\cdots,n_D)$, comprising the measured number of photons in different experimental configurations, {\it e.g.} at different spatial positions or upon illumination with different beams (see Sec.~\ref{sec:minflux_precision}). The goal of the estimation algorithm is to provide estimators $\hat\bxi$ of the true parameters $\bxi$, given the detection outcomes $\bn$. 

The photon counts $\bn$ are random variables and follow the statistical distribution $P(\bn|\bxi,\btheta)$, conditional on the values of all the unknown parameters. These include the target variables $\bxi$ that we want to estimate, as well as additional nuisance parameters $\btheta$ associated to the specific experimental setting. Although these have a limited interest, their values are a priori unkown, thus a multi-parameter estimation approach is followed, considering both
$\bxi$ and $\btheta$. Furthermore, these parameters are often correlated, so that the estimation precision on $\bxi$ is affected by the uncertainty about $\btheta$. 

In light of these reasons, when evaluating the minimum estimation error on $\bxi$, one has to start by computing the CRB for the full vector of parameters $\bpi = (\bxi,\btheta)$: 
\begin{align}\label{eq:crb_theory}
    \cov(\bpi) \geq F(\bpi)^{-1},
\end{align}
where $\cov(\bpi)$ is the covariance matrix of the estimators, with elements
\begin{align}\label{eq:cov_general}
    \cov(\bpi)_{ij} = \langle(\hat\pi_i-\pi_i)(\hat\pi_j-\pi_j)\rangle.
\end{align}
All the expectation values are computed on $P(\bn|\bpi)$, 
whereas $F(\bpi)$ is the Fisher information matrix of the detection statistics (see Sec.~\ref{sec:methods}). For each parameter $\xi_i$, the associated minimum MSE is bounded by:
\begin{align}\label{eq:general_crb_single}
    \sigma^2_i = \cov(\bpi)_{ii} \geq \left(F(\bpi)^{-1}\right)_{ii}.
\end{align}
In single-parameter LET, the nuisance parameters $\btheta$ are considered known, hence the CRB remains in the same form as Eq. \eqref{eq:general_crb_single}, but with the Fisher matrix restricted solely to the block in $F(\bpi)$ pertaining to the parameters $\bxi$. In this case, the correlation between $\bxi$ and $\btheta$ is neglected.

We illustrate this result with a simple example. Consider a single emitter in 1D being stimulated, with all the measured light collected in a single outcome signal, {\it i.e.}, $\bn = n$. The number of photons is a random variable, governed by Poissonian statistics:
\begin{align}\label{eq:Poissonian}
    P(n|\mu) = e^{-\mu}\frac{\mu^{n}}{n!},
\end{align}
where $\mu$ is the mean number of detected photons, that  depends on the position $\bxi = x$ of the emitter, as well as on other factors accounting for the photons generation and collection mechanisms. Without loss of generality, we can group these factors together, considering a single nuisance parameters effective  brightness $\btheta=B$.  
In all the experiments, the brightness is a priori unknown, and we have to consider the joint estimation of the target parameter $x$, in the presence of the nuisance parameter $B$. According to LET~\cite{Lehmann2006}, the Fisher information matrix for the parameter vector $\pi=(x,B)$ is computed as (see Sec.~\ref{sec:methods}) 
\begin{align}
    F(x,B) = \frac{1}{\mu(x,B)} \left(\begin{array}{cc}
    \left(\pd{\mu}{x}\right)^2 & \pd{\mu}{x}\cdot\pd{\mu}{B}\\
    \pd{\mu}{B} \cdot\pd{\mu}{x} & \left(\pd{\mu}{B}\right)^2
    \end{array}\right).
\end{align}
Consequently, the CRB on the MSE on $x$ is
\begin{align}\label{eq:1par_brightness_crb}
    \sigma^2_{1_{\rm MP}} \geq \left(\frac{1}{F(x,B)}\right)_{11} = \left(F(x,B)_{11}-\frac{F(x,B)_{12}\cdot F(x,B)_{21}}{F(x,B)_{22}}\right)^{-1}.
\end{align}
On the other hand, in the single-parameter approach the brightness $B$ is supposed to be known; this amounts to an infinite $F(x,B)_{22}$, so that the CRB  reads directly
\begin{align}\label{eq:1par_brightness_crb_wo_nuisance}
    \sigma_{1_{\rm SP}}^2 \geq \frac{1}{F(x,B)_{11}}
\end{align}
and the error can thus be smaller than the one in \eqref{eq:1par_brightness_crb}, predicted in the presence of the nuisance.

Therefore, in the case that some experimental nuisance parameters are neglected, the prediction of the minimum estimation may lead to an overly optimistic CRB. The difference between the two errors \eqref{eq:1par_brightness_crb} and \eqref{eq:1par_brightness_crb_wo_nuisance} is small when the nuisance parameters are weakly correlated with the true parameters,{\it i.e.}  when $F(x,B)_{12}$ is small. However, the photon detection statistics in OFM, and in particular its mean value $\mu$, depend on both target and nuisance parameters, and their statistical independence is unlikely.

\subsection{Emitter localisation precision in MINFLUX microscopy}\label{sec:minflux_precision} 

We consider the MINFLUX microscopy configuration, where the position of the emitter $\bxi$ is evaluated using multiple illuminating beams, with intensities $I(\bxi - \br_i)$. Here, $\br_i$ denotes the coordinates of the beam center, at distance $\frac{L}{2}$ from the origin, {\it i.e.}, $|\br_i|=\frac L2$ for all $i$. Note that the illuminating beams have an intensity minimum at their center, rather than a maximum, as in conventional OFM approaches.  If the emitter is located exactly at the center of either beam, it is not excited, thus it does not emit photons; conversely, if the emitter is slightly shifted from the beam center, we  expect a small photon count rate. Collecting a sufficiently large number of photons facilitates the emitter localisation and decreases the estimation error on $\bxi$, as reported by~\cite{Balzarotti2017,Gwosch2020}. After illumination with the $i$-th beam, the outcome $n_i$ is the total number of photons detected in the entire observed region, following the Poissonian statistics $P(n_i|\mu_i)$  in \eqref{eq:Poissonian}, with a mean value given by
\begin{align}
    \mu_i(\bxi,B) = B \cdot I(\bxi-\br_i),
\end{align}
where $B$ is the unknown brightness parameter. Since the positions $\br_i$ are known, $B$ represents the only nuisance parameter, and we study its role within LET. 

\subsubsection{1D localisation}
We start by considering a 1D localisation problem $\bxi = x$, with two illuminating beams centered symmetrically around the origin at a relative distance $L$: $\br_1 = -\frac L2$ and $\br_2 = \frac L2$. Furthermore, we restrict our attention to a quadratic beam shape,
\begin{align}
    I(x) = \frac{x^2}{s^2},
\end{align}
where $s$ is the beam width; this is a satisfactory  approximation if $|x|\lesssim s$. 
According to the mathematical framework presented in Sec.~\ref{sec:general} (see also Sec.~\ref{sec:methods}), we obtain from the photodetection outcomes $\bn=(n_1,n_2)$ a CRB on the minimum uncertainty in the estimation of $x$ writing:
\begin{align}\label{eq:minflux_1d_crb_true}
    \sigma^2_{1_{\rm MP}} \geq \frac{s^2 \left(L^2+4 x^2\right)}{8 B L^2}.
\end{align}
We can compare this bound to the single-parameter estimation  derived in~\cite{Balzarotti2017}: there, the authors keep the number of total events $N=n_1+n_2$ fixed. Under this condition, the corresponding CRB is
\begin{align}\label{eq:minflux_1d_crb_balzarotti}
    \sigma^2_{1_{\rm SP}} \geq \frac{\left(L^2+4 x^2\right)^2}{16 N L^2}.
\end{align}
The two approaches  provide different localisation errors. In particular, for an emitter placed at the origin $x=0$, the precision in \eqref{eq:minflux_1d_crb_balzarotti}, obtained neglecting the nuisance parameter, decreases quadratically with the beam separation $L$, whereas it is independent of $L$ in \eqref{eq:minflux_1d_crb_true}. 
Here, the single-parameter framework considers the detected photon number $N$ fixed and independent of $L$. On the other hand, the average photon number detected in the multi-parameter measurement is 
\begin{equation}\label{eq:mean_photon_number}
    \bar N = \mu_1(0,B)+\mu_2(0,B) = \frac{B L^2}{2 s^2},
\end{equation}
which decreases quadratically with the beam separation $L$. If we identify $N=\bar N$, and substitute \eqref{eq:mean_photon_number} into the single-parameter error \eqref{eq:minflux_1d_crb_balzarotti}, we recover the multi-parameter error \eqref{eq:minflux_1d_crb_true}.

Therefore, the single-parameter LET predicts that reducing the beam separation $L$ may indeed enhance the localisation precision, whereas the multi-parameter treatment accounts for the consequent drop of the average number of detected photons. The proposed multi-parameter framework unveils this trade-off, and makes it possible to compare different configurations for different $L$ values, under the same experimental constraints. 
In fact, in ~\cite{Balzarotti2017}, the total number of photons is kept fixed for every value of $L$, implying that the reduction in the emission intensity, that occurs for small values of $L$, is compensated, for instance, by larger beam intensities of longer exposure times. 
Instead,  multi-parameter LET includes all these effects in the nuisance brightness parameter $B$, that, for a generic $x$, 
can be estimated with a precision satisfying  
\begin{align}\label{eq:minflux_1d_B_crb_true}
    \sigma_{2_{\rm MP}}^2 \geq \frac{2 B s^2}{L^2}.
\end{align}

Furthermore, the multi-parameter framework of \eqref{eq:minflux_1d_crb_true} uncovers another key property: the estimation error on $x$ depends also on the beam width $s$, which is often related to the illuminating wavelength.
Therefore,  assuming that one is able to control the activation of a single emitter within a region of size $L$, its localisation precision is still influenced by the illuminating wavelength, analogously to the diffraction limit. Clearly, super-resolution can be obtained only by increasing the brightness. 

\begin{figure}[H]
    \centering
    \includegraphics[width=\textwidth]{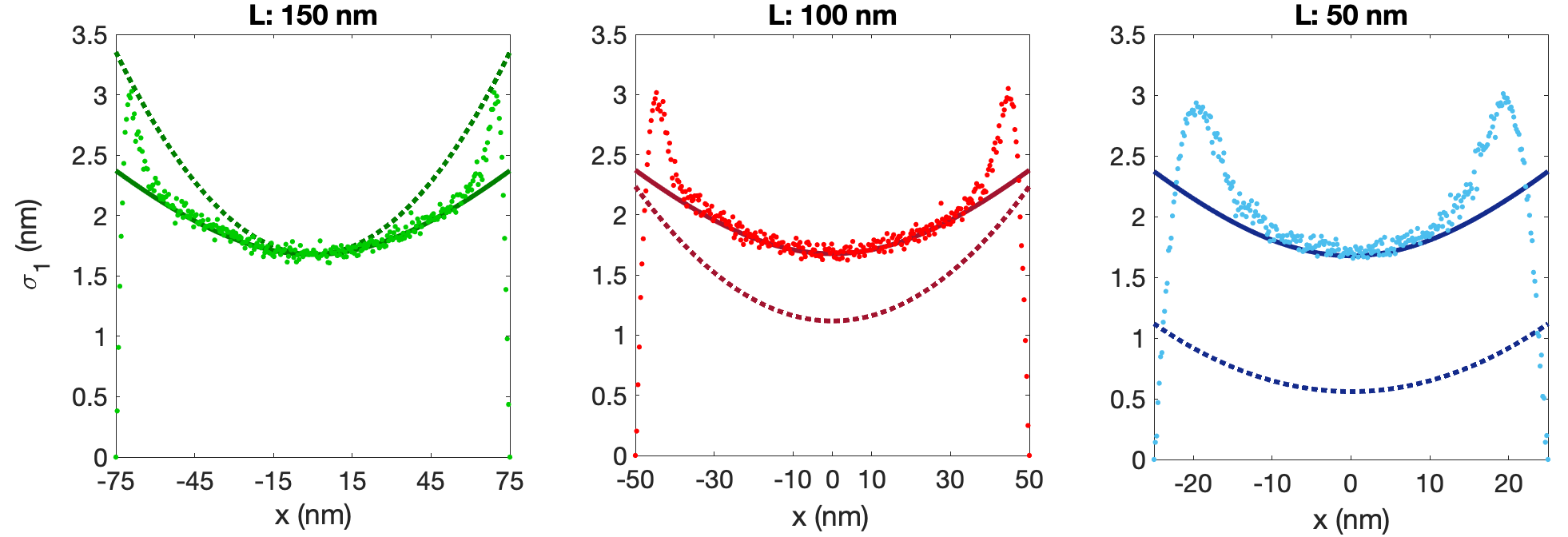}
    \caption{Analytical and numerical precision for the 1D localisation of an emitter at position $x$, for  $L=150nm$, $L=100nm$, and $L=50nm$ and $B=4000$. Solid lines correspond to the multi-parameter CRB \eqref{eq:minflux_1d_crb_true}, and dashed ones to the single-parameter CRB \eqref{eq:minflux_1d_crb_balzarotti}. Dots have been obtained from numerical simulations over $1000$ repetitions of the experiment.}
    \label{fig:crb_1d_num}
\end{figure}

Quantitative considerations can be also drawn from an inspection of Fig.~\ref{fig:crb_1d_num}, where we compare the square-root of the two error bounds \eqref{eq:minflux_1d_crb_true} and \eqref{eq:minflux_1d_crb_balzarotti} for three values of $L$, as a function of the emitter position. We fix $B$ so that, if $x=0$ the two values $N$ and $\bar N$ \eqref{eq:mean_photon_number} coincide for $L=150nm$. A comparison of the two analytical expressions reveals that in the multi-parameter estimation, the error is constant at the origin for different values of $L$, and this does not occur in the single-parameter case.  
Moreover, as the emitter is displaced from the origin,  the associated error is expected to increase less with the multi-parameter approach, regardless of $L$.

We compare the theoretical predictions with the results of numerical simulations for an OFM localisation experiment (see Sec.~\ref{sec:methods}). These were obtained by extracting photon counts according to the Poisson statistics and employing the maximum-likelihood estimators
\begin{align}\label{eq:estimators_n1neqn2}
    \hat x = \frac{L \left(\sqrt{n_1}-\sqrt{n_2}\right)}{2 \left(\sqrt{n_1}+\sqrt{n_2}\right)}, \quad \hat B = \frac{s^2 \left(\sqrt{n_1}+\sqrt{n_2}\right)^2}{L^2}.
\end{align}
Repeating this single experiment for several runs, one is able to calculate the root MSE of the estimates for each possible emitter position. 
The maximum-likelihood estimator for the position 
coincides with that of~\cite{Balzarotti2017} however, given the different error analysis, the MSE is expected to be different.

From an inspection of Fig.~\ref{fig:crb_1d_num}, it is evident that the numerical results accurately follow the multi-parameter LET. However, there is a significant discrepancy  when the emitter is close to the centre of one of the two illumination beams, {\it i.e.} for large values of $x$. In this situation the estimation becomes particularly challenging, due to the very small photon counts expected from one of the beams, whose fluctuations can more heavily affect the convergence of the maximum-likelihood estimator. However, even though these operating conditions are quite uncommon, the issue can be partly mitigated by increasing the number of repetitions of the experiment.

\subsubsection{2D localisation}
We now extend the analysis to a 2D localisation problem, {\it i.e.}, $\bxi = (x,y)$, and consider four illuminating beams: one is centered at the origin, $\br_{0} = (0,0)$, and the other three placed symmetrically on a circumference of radius $\frac L2$:
\begin{align}\label{eq:2d_beam_positions_0}
    &\br_0 = (0, 0),\\
    &\br_1 = \frac L2 (1, 0),\\
    &\br_2 = \frac L2\left(\cos\frac{2\pi}{3}, \sin\frac{2\pi}{3}\right) = \frac L2 \left(-\frac12,\frac{\sqrt3}2\right),\\
    &\br_3 = \frac L2 \left(\cos\frac{4\pi}{3}, \sin\frac{4\pi}{3}\right) = \frac L2 \left(-\frac12,-\frac{\sqrt3}2\right).\label{eq:2d_beam_positions_3}
\end{align}
As in the 1D case, we restrict our attention to paraboloid-shaped beams:
\begin{align}
    I(\bxi-\br_i) = \frac{|\bxi-\br_i|^2}{s^2}.
\end{align}

\begin{figure}[H]
    \centering
    \includegraphics[width=\textwidth]{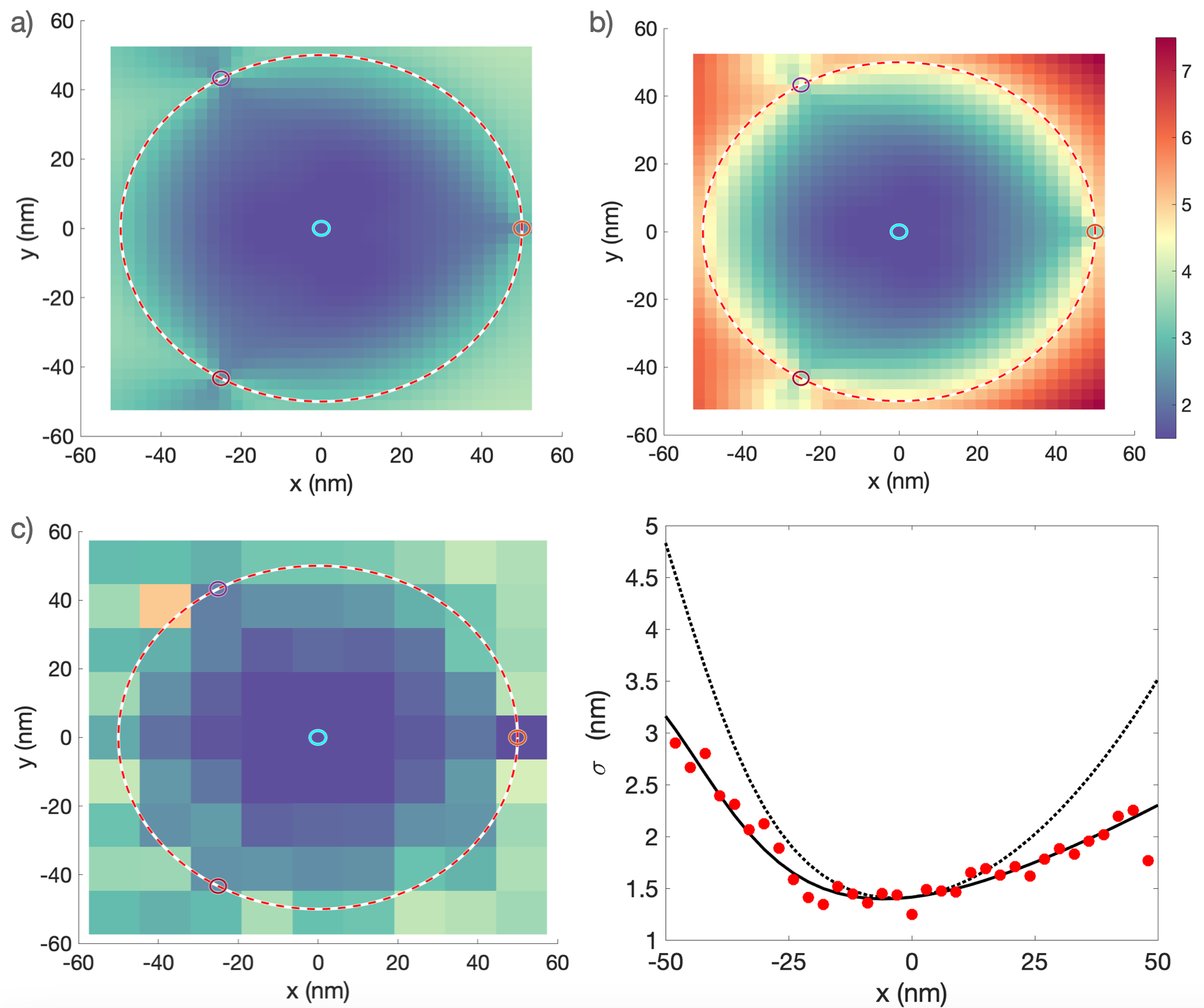}
    \caption{2D localisation results. a)  Density plots ($35\times35$) of the analytical error for the localisation of the emitter at position $(x,y)$ a) multiparameter CRB (Eq.~\eqref{eq:minflux_2d_crb_true})  b) single-parameter CRB obtained by fixing the total number of detected photons $N=500$, Eq.~\eqref{eq:minflux_2d_crb_balzarotti}.
    c) Density plot ($9\times9$) of the numerical error.  $L=100nm$ and the brightness is set $B=6000$ so that the errors at the origin coincide in the two approaches. The dashed line corresponds to the circumference of radius $\frac{L}{2}$, and the circles correspond to the position of the four beams  (\ref{eq:2d_beam_positions_0}-\ref{eq:2d_beam_positions_3}), cyan, orange, purple, and red respectively. 
    d) cut of the 2D profiles for $y=0$: multi-parameter CRB (solid line), single-parameter CRB (dashed line) and plot of numerical simulations (dots).}
    \label{fig:crb_2d}
\end{figure}

We now aim at estimating the two coordinates $(x,y)$, with the brightness $B$ acting as a nuisance parameter.
According to the mathematical framework presented in Sec.~\ref{sec:general} (see also Sec.~\ref{sec:methods}), we can bound the average mean square error on the estimation of $(x,y)$ as 
\begin{align}\label{eq:minflux_2d_crb_true}
\bar \sigma^2_{{\rm MP}}& = \frac{\sigma^2_{1_{\rm MP}}+\sigma^2_{2_{\rm MP}}}{2}\\ 
    &\geq \frac{2 s^2 \left(x^2+y^2\right) \left(L^6-16 L^3
   \left(x^3-3 x y^2\right)+64
   \left(x^2+y^2\right)^3\right)}{3 B L^2 \left(5
   L^4 \left(x^2+y^2\right)-4 L^3 \left(x^3-3 x
   y^2\right)-28 L^2 \left(x^2+y^2\right)^2+64
   \left(x^2+y^2\right)^3\right)},
\end{align}
In particular, for an emitter placed at the origin $(x,y)=(0,0)$, the multi-parameter estimation provides an average uncertainty bounded by
\begin{align}\label{eq:minflux_2d_crb_xi0_true}
     \bar \sigma^2_{{\rm MP}}\Bigg|_{\bxi = 0} \geq \frac{2s^2}{15B}.
\end{align}
In contrast, in single-parameter estimation~\cite{Balzarotti2017}, the same quantity computed by fixing the total number of detected photons $N = n_0+n_1+n_2+n_3$ is: 
\begin{align}\label{eq:minflux_2d_crb_xi0_balzarotti}
    \bar \sigma^2_{{\rm SP}}\Bigg|_{\bxi = 0} \geq \frac{L^2}{10 N}. 
\end{align}
Therefore, also in the 2D case, the error does not depend on $L^2$ when the brightness is included in the LET.
\\
Fig.~\ref{fig:crb_2d} a,b show the error predictions \eqref{eq:minflux_1d_crb_true} and \eqref{eq:minflux_1d_crb_balzarotti} via density plots, as a function of the emitter's 2D position for $L=100 nm$ and brightness $B=\frac{4 N s^2}{3 L^2}$, chosen analogously to the 1D case. The predicted error is generally smaller in multi-parameter LET. 
In addition, there are areas in the periphery, close to the beam centers, where the localisation is more difficult, because those pixels are at equal distance from the other two beams, and the expected photon counts are approximately the same for the two illuminations.

Also in this case we performed extensive numerical simulations of the OFM localisation experiment (Fig.~\ref{fig:crb_2d} c).  However, because the maximum likelihood estimator cannot be analytically computed, we used numerical calculations (see Sec.~\ref{sec:methods}), and due to the required CPU time we had to reduce the number of pixels.  To confirm the validity of our findings, we report in Fig.~\ref{fig:crb_2d}  d  a 1D-cut performed at a higher resolution in the $x$ direction, with fixed (but still unknown) $y$. 

\subsection{Emitter localisation precision in STED microscopy}
The general framework introduced in Sec.~\ref{sec:general} can be applied to any OFM technique, as long as the relevant nuisance parameters are correctly identified, thus allowing for a fair comparison between different approaches. Here, we compare two scenarios: MINFLUX and STED microscopy, restricting to a 1D case for the sake of simplicity.

In the STED microscope, the illuminating beam scans a field of view (FOV) of total area $L$ centered at the origin, and it is focused on a region of width $\Delta=L/m$, where $m$ is total number of scans. The point illumination is a Gaussian maximum-flux beam, of known width $\sigma$ and unknown brightness $B_{\rm sted}$. In order to make a fair comparison with MINFLUX, we set $B_{\rm sted} = \frac{2 B}{m}$, so that the total brightness is the same for both techniques.
After the $i$-th point illumination, $n_i\,  (i=1,2,..,m)$  photons are detected, following a Poissonian statistics of mean:
\begin{align}\label{eq:mu_confocal}
    \mu_i^{\rm sted} = \frac{B_{\rm sted}}{\sqrt{2\pi \sigma}} e^{-\frac{(x-x_i)^2}{2\sigma}}.
\end{align}
The resulting FIM and CRB for the parameters $x$ and $B_{\rm sted}$ can be obtained along the general lines of Sec.~\ref{sec:general} (see Sec.~\ref{sec:methods} for a detailed description), opting for a numerical evaluation. 

\begin{figure}[H]
    \centering
    \includegraphics[width=\textwidth]{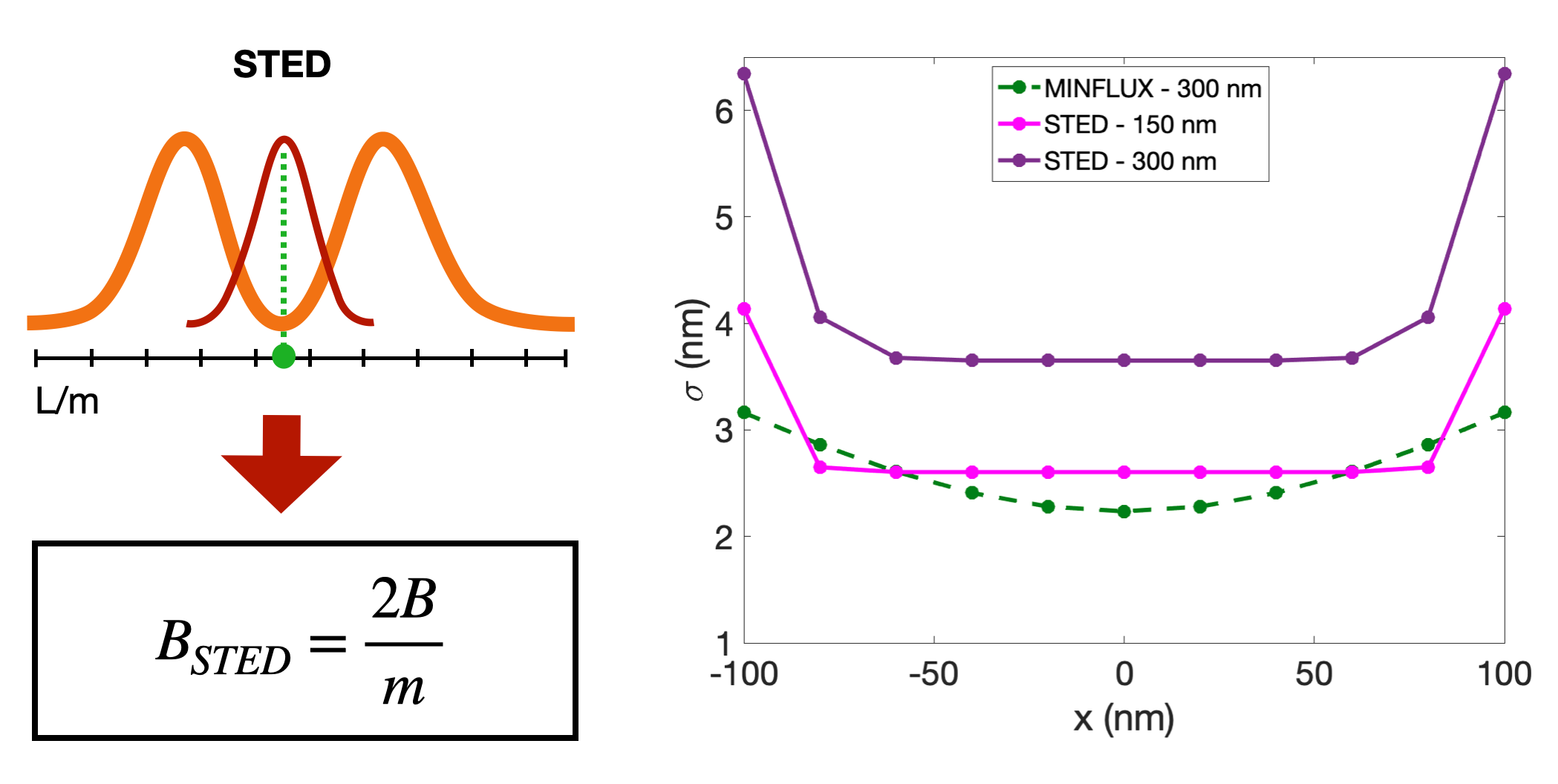}
    \caption{Minimum root MSE for the localisation of an emitter at position $x$ with MINFLUX and STED microscopes. The field of view is $L=200nm$.  The brightness of a single MINFLUX beam is $B=2500$ and for STED is $B_{\rm sted} = 150$, considering $m=30$ pixels. The beam width are $\sigma=150$ and $300nm$}
    \label{fig:confocal}
\end{figure}

Fig.~\ref{fig:confocal} reports the minimum localisation error of an emitter in a 1D configuration for STED and MINFLUX microscopy, leveraging our multi-parameter analysis. The size of FOV is  $L=200nm$ and we use beams of width $s=300nm$ for MINFLUX and two different widths $\sigma = 300nm, 150nm$ for STED microscopy; the pixel size is $20nm$. In the case of  STED,$\sigma$ is the point spread function waist \cite{Cerutti2021}.
We observe that the STED microscope exhibits an error plateau due to the scanning process. 
Furthermore, the plot shows that MINFLUX presents a smaller localisation error than STED for the same value of beam width $s=\sigma=300nm$, and this value is quite comparable when we reduce the STED point spread function width to $\sigma=150nm$ .  

\section{Discussion}\label{sec:discussion}
A wide array of super-resolution OFM techniques have been proposed in literature, each reporting some advancements in image quality, contrast  or resolution. In this work, we move beyond the image parameter evaluation, and introduce a rigorous mathematical quantifier of the localisation error. 
Our method allows to identify the experimental parameters that affect the precision of any OFM technique, and take them into account in the error analysis. In turn, this makes it possible to analyse the precision of each technique as a function of its parameters, as well as to compare different techniques on an equal footing. 

We established the multi-parameter LET as the tool of choice for the precision analysis of OFM experiments. We employed this method to address a crucial issue in previous applications~\cite{Chao2016,Balzarotti2017,Ober2020,Kalisvaart2022,Zhan2022,Xiu2023}, namely the presence of additional parameters that the experimenter is not interested in estimating,  but that are nonetheless a priori unknown and affect the final precision.

As a key example, we discussed the case of the effective brightness, an unknown parameter that influences the mean number of photons detected in an OFM experiment. We applied our method to calculate the localisation error of MINFLUX and STED microscopy. 
Leveraging multi-parameter LET, we shed light on the trade-off between beam separation $L$ and total number of detected photons, showing that the precision of MINFLUX microscopy cannot be increased by only decreasing $L$. 
Thanks to a compact parametrisation, the relative merits and convenience of different experimental options can be compared without ambiguities, as we illustrated for STED microscopy. 

In light of these results, we anticipate that our method will advance the understanding of super-resolution microscopy precision limits. 
Our work therefore provides a starting point to investigate a variety of other critical points in OFM. In particular, we identify the following  questions for prospective future case studies:
 (i) establishing the fundamental resolution limits of OFM techniques, {\it i.e.}, tackling the problem of resolving multiple fluorescent emitters; (ii) introducing further nuisance parameters that influence OFM techniques, achieving a precise evaluation of their effect on the final error; (iii) determining the precision limits of more complex super-resolution techniques, {\it e.g.}, those based on stochastic fluorophore activation. 
The introduction of this new tool will consolidate further explorations in quantitative life sciences.

\section{Methods}\label{sec:methods}

\subsection{Short summary of LET}
When trying to estimate the true values of parameters $\bpi$ from detection outcomes $\bn$, one employs estimators $\hat\bpi$ that are function of the outcomes themselves and of their probability distribution $P(\bn|\bpi)$. The covariance matrix \eqref{eq:cov_general} summarizes the mean error and correlation of the estimators $\hat\bpi$. For a fixed detection statistics, the covariance matrix of any unbiased\footnote{An unbiased estimator is one whose expectation reproduces the true value, {\it i.e.}, $\langle\hat\bpi\rangle = \bpi$. The maximum-likelihood estimator is asymptotically unbiased, {\it i.e.}, it is unbiased in the limit of a large number of repetitions of the experiment.} estimator $\hat\bpi$ can be bounded by the CRB \eqref{eq:crb_theory}. The latter is computed in terms of the Fisher information matrix (FIM), whose elements are
\begin{align}\label{eq:fim}
    F(\bpi)_{ij}=\left\langle\left(\partial_{\pi_i}\log P(\bn|\bpi)\right)\left(\partial_{\pi_j}\log P(\bn|\bpi)\right)\right\rangle,
\end{align}
where the expectation is taken with respect to the probability distribution of the outcomes, conditional on the true parameters, {\it i.e.}, $P(\bn|\bpi)$.

Depending on the form of $P(\bn|\bpi)$, it can be convenient to first evaluate the FIM for a simpler set of parameters $\bpi'$ and then change variables to the final set $\bpi$. This is done via the Jacobian matrix of elements
\begin{align}
    J_{ij} = \pd{\pi'_i(\bpi)}{{\pi_j}},
\end{align}
applying the transformation
\begin{align}
    F(\bpi) = J^T F(\bpi') J. 
\end{align}

Once the CRB on the estimation error is computed, one is usually interested in finding an actual estimator $\hat \bpi$ that attains the bound. In the case of multi-parameter estimation, it is important to note that in general the CRB \eqref{eq:crb_theory} is not attainable for all the parameters $\bpi$ simultaneously, {\it i.e.}, there might not exist an estimator that attains the bounds
\begin{align}\label{eq:multi_crb}
    \sigma^2_i \geq \frac{1}{F(\bpi)_{ii}}
\end{align}
for all $\pi_i$ simultaneously, unless the corresponding parameters are uncorrelated. For example, the bound for $\pi_i$ and $\pi_j$ can be attained with the same estimator if $F(\bpi)_{ij}=F(\bpi)_{ji}=0$. 

The maximum-likelihood estimator (MLE) is an estimator that asymptotically attains the CRB in the limit of a large number of repetitions. In practice, this means that, in order to obtain a mean square error close to the CRB, one needs to run the experiment several times, each time computing an estimate of the true parameter, and then average all such estimates. The obtained average estimate then has a mean square error close to the CRB, the larger the number of repetitions is. 

The expression of the MLE can be computed as a function of the detection outcomes $\bn$ by maximizing the likelihood function
\begin{align}\label{eq:likelihood}
    \cL(\bpi) = P(\bn|\bpi),
\end{align}
which, assuming that all values $\bpi$ are equiprobable in a small interval (hence the adjective local), is proportional to the probability that certain values of the parameters might have produced those specific values of the detection outcomes.
In practice, one needs to set the first derivatives of \eqref{eq:likelihood} (or of its logarithm) to zero, solve the corresponding equations and then choose the solution that gives the largest value of \eqref{eq:likelihood}. In this way, one identifies the estimates $\hat\bpi(\bn)$ as those values of the parameters that are most likely to have produced the outcomes $\bn$. 

Furthermore, if the problem presents additional parameters, {\it i.e.}, if the statistics $P(\bn|\bpi)$ depends on a set of parameters $\bpi=(\bxi,\btheta)$, comprising the parameters $\bxi$ we want to estimate plus so-called nuisance parameters $\btheta$, the estimation becomes more difficult. 
Indeed, either it is possible to assign known and fixed values to the nuisance parameters $\btheta$, so that they effectively become constants in the calculation of the FIM, or else they are unknown as much as the $\bxi$ and hence need to be estimated from the detection outcomes too. 
Clearly, in the latter case, the only hope of decoupling the estimation of $\bxi$ from that of $\btheta$ relies on their being uncorrelated, which is generally not the case. 

In conclusion, when nuisance parameters are involved, even the estimation of a single quantity becomes a multi-dimensional estimation problem treatable only with the full FIM \eqref{eq:fim}, giving rise to corrections of the precision, as illustrated in Sec.~\ref{sec:general}.

\subsection{Calculation of the CRB and MLE for MINFLUX microscopy}
In the 1D case we illuminate with  parabolic beams, inducing a mean number of detected photons
\begin{align}\label{eq:mean_1d}
    \mu_i = \frac{B}{s^2}(x-x_i)^2,
\end{align}
where $x_1 = -\frac L2$ and $x_2 = \frac L2$. The photon detection statistics for each beam is thus independent and Poissonian:
\begin{align}\label{eq:statistics_1d}
    P(n_1,n_2|x,B) = P(n_1|x,B) \cdot P(n_2|x,B) = \prod_{i=1}^2 e^{-\mu_i}\frac{\mu_i^{n_i}}{n_i!}.
\end{align}
We want to estimate the parameters $x$ and $B$ but, given the product form \eqref{eq:statistics_1d}, it is easier to start from the estimation of $\mu_1$ and $\mu_2$, which turn out to be uncorrelated. Therefore, their FIM is
\begin{align}
    F(\mu_1,\mu_2) = \left(\begin{array}{cc}
        \frac1{\mu_1} & 0 \\
        0 & \frac1{\mu_2}
    \end{array}\right).
\end{align}

Then we change variables to the parameters $(x,B)$ via the transformation \eqref{eq:mean_1d}.
Accordingly, the Jacobian for this change of variable is
\begin{align}
    J = \frac1{4s^2}\left(
\begin{array}{cc}
 4B (2 x +L) & \left(2x
   +L\right)^2 \\
4B (2 x -L) & \left(2x
   -L\right)^2 \\
\end{array}
\right).
\end{align}
Hence the FIM for the final parameters is
\begin{align}\label{eq:fmi_minflux_ult_dd}
    F(x,B) = \frac1{s^2}\left(
\begin{array}{cc}
 8 B & 4 x  \\
 4 x & \frac{L^2+4 x ^2}{2 B} \\
\end{array}
\right).
\end{align}

By applying \eqref{eq:1par_brightness_crb}, we conclude that the minimum mean square error attainable on the estimation of $x$ is given by \eqref{eq:minflux_1d_crb_true} and similarly for the estimation of $B$ by \eqref{eq:minflux_1d_B_crb_true}. 
Furthermore, from the FIM above it is evident that the two parameters $x$ and $B$ are correlated, the more so the farther the emitter is situated from the origin.

Also the maximum-likelihood estimators \eqref{eq:estimators_n1neqn2} can be computed exactly in this case. The likelihood function is simply $\cL(n_1,n_2) = P(n_1,n_2|x,B)$, given by \eqref{eq:statistics_1d} and, given $n_1, n_2$, we look for values $x,B$ satisfying 
\begin{align}
    \pd{\cL}{x} = 0,\quad \pd{\cL}{B}=0.
\end{align}
The second equation has trivial solutions $B=0,\infty$, where the function is clearly minimum at fixed $x$, and non-trivial solution
\begin{align}
    B = \frac{2 s^2 (n_1+n_2)}{L^2+4 x ^2}.
\end{align}
Plugging this into the first equation we obtain the non-trivial solutions
\begin{align}\label{eq:estimators_n1neqn2_gen}
    \hat x_{\pm} = \frac{L \left(\sqrt{n_1}\pm\sqrt{n_2}\right)}{2 \left(\sqrt{n_1}\mp\sqrt{n_2}\right)}, \quad \hat B_\pm = \frac{s^2 \left(\sqrt{n_1}\mp\sqrt{n_2}\right)^2}{L^2},
\end{align}
of which only the $(\hat x_-, \hat B_-)$ is significant since $|\hat x_+|>\frac L2$ predicts an emitter position outside of the range where the quadratic approximation of the beam holds. \\

Similar calculations can be carried out for the 2D case. For simplicity, we still consider the case of parabolic beams with mean number of detected photons
\begin{align}
    \mu_i = \frac{B}{s^2}|\bxi - \br_i|^2 = \frac{B}{s^2} \left[(x-x_i)^2+(y-y_i)^2\right],
\end{align}
where $\br_i = (x_i,y_i)$ is the 2D position of the center of the $i$-th beam with respect to the origin for $i=0,\cdots,3$, as given in Eqs. (\ref{eq:2d_beam_positions_0}-\ref{eq:2d_beam_positions_3}). The photon detection statistics for each beam is still independent and Poissonian, $P(\bn|x,y,B) = \prod_{i=0}^3 P(n_i|x,y,B)$. Starting again from the estimation of $\mu_i$ we have
\begin{align}
    F(\mu_0,\mu_1,\mu_2,\mu_3) = \left(\begin{array}{cccc}
        \frac1{\mu_0} & 0 & 0 & 0\\
        0 & \frac1{\mu_1} & 0 & 0 \\
        0 & 0 & \frac1{\mu_2} & 0 \\
        0 & 0 & 0 & \frac1{\mu_3} 
    \end{array}\right).
\end{align}
In order to change variables to the parameters $x,y,B$, {\it i.e.}, the emitter's 2D coordinates and the nuisance parameter $B$, we compute the Jacobian
\begin{align}
J=\frac1{4s^2}\left(
\begin{array}{ccc}
 8 B x & 8 B y & 4
   \left(x^2+y^2\right) \\
 -4 B (L-2 x) & 8 B y & (L-2
   x)^2+4 y^2 \\
 2 B (L+4 x) & B \left(8 y-2
   \sqrt{3} L\right) & \frac{
   \left(L+4x\right
   )^2+\left(4y-\sqrt{3}
   L\right)^2}{4} \\
 2 B (L+4 x) & 2 B \left(\sqrt{3}
   L+4 y\right) & 
   \frac{\left(L+4x\right
   )^2+\left(\sqrt{3}
   L+4y\right)^2}{4} \\
\end{array}
\right).
\end{align}
The resulting FIM is
\begin{align}
    F(x,y,B) = \frac1{4s^2} \left(\begin{array}{ccc}
        F_{11} & F_{12} & 32x \\
        F_{21} & F_{22} & 32y \\
        32x & 32y & \frac{3 L^2+16
   \left(x^2+y^2\right)}{B} 
    \end{array}\right),
\end{align}
where we omit the explicit expression of the submatrix $F_{i,j}$, due to its length. From the FIM one can obtain the average mean square error on the estimation of $x$ and $y$, \eqref{eq:minflux_2d_crb_true}, by taking the inverse of $F(x,y,B)$ and averaging the first two diagonal entries. 
As for the maximum likelihood estimators, in this case it is not possible to provide a fully analytical solution. Indeed, the likelihood $\cL(\bn) = P(\bn|x,y,B)$ has a stationary point in $B$ for 
\begin{align}\label{eq:estimator_b_2d}
    B = \frac{4 s^2 (n_0+n_1+n_2+n_3)}{3 L^2+16 x^2+16
   y^2},
\end{align}
but unfortunately the stationarity equations for $x$ and $y$, 
\begin{align}\label{eq:stat_2d}
    \pd{\cL}{x} = 0, \quad \pd{\cL}{y} = 0,
\end{align}
cannot be solved analytically in closed form, even after substituting the optimal value \eqref{eq:estimator_b_2d}.
Therefore, in practice one solves these equations numerically for the given values of $\bn$ and then identifies the optimal estimator wtih the solution that provides the largest value of the likelihood.  

Finally, for completeness we report here the full expression of the CRB for 2D localisation using the method of Ref.~\cite{Balzarotti2017}, {\it i.e.}, fixing the total number of detected photons:
\begin{align}
    \label{eq:minflux_2d_crb_balzarotti}
    \bar \sigma_{\rm bal}^2 &= \left(x^2+y^2\right) \left((L-2 x)^2+4 y^2\right) \\
   &\cdot\frac{\left(-4
   y^2 \left(L^2-4 L x-8 x^2\right)+\left(L^2+2 L x+4
   x^2\right)^2+16 y^4\right)\left(3 L^2+16
   \left(x^2+y^2\right)\right)}
{6 L^2 N \left(5 L^4
   \left(x^2+y^2\right)-4 L^3 \left(x^3-3 x y^2\right)-28 L^2
   \left(x^2+y^2\right)^2+64 \left(x^2+y^2\right)^3\right)} \nonumber
\end{align}
which, in particular, for an emitter at the origin gives \eqref{eq:minflux_2d_crb_xi0_balzarotti}.

\subsection{Calculation of the CRB for STED microscopy}
The estimation error can be evaluated starting from the FIM for the independent parameters \eqref{eq:mu_confocal}: $F(\bmu^{\rm sted}) = {\rm diag}(\frac{1}{\mu_1^{\rm sted}},\cdots,\frac{1}{\mu_m^{\rm sted}})$. The Jacobian for the change of variables to $x, B_{\rm sted}$ is a $m\times 2$ rectangular matrix with rows of elements
\begin{align}
    J_{i,1} &= \frac{B (x-x_i ) e^{-\frac{(x-x_i )^2}{2 \sigma}}}{\sqrt{2 \pi }\sigma^{3/2}},\\
    J_{i,2} &= \frac{\mu_i^{\rm sted}}{B_{\rm sted}},
\end{align}
for all $i=1,\cdots,m$.
The final FIM can be large for a large numebr of pixels, hence we resort to numerical evaluation of its inverse. 

\subsection{Numerical methods}
We performed numerical simulations of the fluorophore localisation experiment with a MINFLUX microscope.
For a given emitter position $\bxi$, the statistics $P(n_i|\bxi,B)$ is completely determined by fixing the position of each illuminating beam, $\br_i$, and the value of the brightness. Once these quantities are fixed, we can simulate a single run of the experiment by generating random values of the photon counts $\bn$ according to the independent Poissonian distribution \eqref{eq:statistics_1d} for 1D and similarly for 2D. 

The obtained photon counts $\bn$ correspond to those detected in a run of the experiment with a certain illumination time. Therefore, the maximum-likelihood estimator $\hat\bxi$ can be computed by plugging the obtained $\bn$ values in the expression \eqref{eq:estimators_n1neqn2} for 1D or by solving the stationariety equations \eqref{eq:stat_2d} in 2D.
One can then repeat this procedure $M$ times, each running a different experiment with detection outcomes $\bn^{(m)}$ and obtaining an estimate $\hat\bxi^{(m)}$ of the true position, for $m=1,\cdots,M$. The mean  
and mean square error of such estimates with respect to the runs provide the final estimate of the position and its variance, which can be compared with the theoretical CRB. 




\backmatter

\bmhead{Acknowledgments}
MR acknowledges support from the PNRR project 816000-2022-SQUID - CUP F83C22002390007 (Young Researchers).
IG and MB are supported by the FET-OPEN-RIA project STORMYTUNE (Grant Agreement number 899587)

\end{document}